\documentclass[fleqn,10pt]{wlscirep}
\usepackage[utf8]{inputenc}
\usepackage[T1]{fontenc}
\usepackage{graphicx}
\usepackage{fontawesome}
\usepackage{hyperref}

\title{Online quantum time series processing with random oscillator networks}

\author[1,*]{Johannes Nokkala}
\affil[1]{Turku Centre for Quantum Physics,  Department for Physics and Astronomy, University of Turku, FI-20014, Turun Yliopisto, Finland}
\affil[1]{IFISC, Instituto de F\'{\i}sica Interdisciplinar y Sistemas Complejos (UIB-CSIC) UIB Campus, E-07122 Palma de Mallorca, Spain}

\affil[*]{jsinok@utu.fi}

\begin{abstract}
Reservoir computing is a powerful machine learning paradigm for online time series processing. It has reached state-of-the-art performance in tasks such as chaotic time series prediction and continuous speech recognition thanks to its unique combination of high computational power and low training cost which sets it aside from alternatives such as traditionally trained recurrent neural networks, and furthermore is amenable to implementations in dedicated hardware, potentially leading to extremely compact and efficient reservoir computers. Recently the use of random quantum systems has been proposed, leveraging the complexity of quantum dynamics for classical time series processing. Extracting the output from a quantum system without disturbing its state too much is problematic however, and can be expected to become a bottleneck in such approaches. Here we propose a reservoir computing inspired approach to online processing of time series consisting of quantum information, sidestepping the measurement problem. We illustrate its power by generalizing two paradigmatic benchmark tasks from classical reservoir computing to quantum information and introducing a task without a classical analogue where a random system is trained to both create and distribute entanglement between systems that never directly interact. Finally, we discuss partial generalizations where only the input or only the output time series is quantum.
\end{abstract}
\begin{document}

\flushbottom
\maketitle

\thispagestyle{empty}

\section*{Introduction}

Tasks where one time series need to be transformed into another include time series forecasting\cite{jaeger2001echo,jaeger2002adaptive}, pattern generation\cite{jaeger2002tutorial,hauser2012role} and pattern recognition\cite{verstraeten2005isolated,soh2012iterative,paquot2012optoelectronic,jalalvand2015real}. In online time series processing both the given data and desired transformed data are functions of time, which separates it from approaches such as first recording the data and later processing it. Instead, the objective is to realize the time dependent function which for a given timestep and input time series up to that step returns the corresponding element of the output time series. Such tasks are also known as temporal tasks. When successful, online time series processing facilitates, e.g., the processing of arbitrarily long sequences of data since the inputs are continuously processed into outputs. This is possible in particular for tasks that can be solved by so called fading memory functions, which are functions well approximated by continuous functions of only a finite number of past inputs\cite{boyd1985fading}. Under typically mild conditions the input time series can be used to drive random dynamical systems such that their internal variables become such fading memory functions, which can then be combined to approximate the desired ouput by training a simple, even linear readout function. This is known as reservoir computing (RC), which is a powerful approach to solving temporal tasks thanks to a remarkably low training cost\cite{lukovsevivcius2012practical,butcher2013reservoir} combined with state-of-the-art performance\cite{schrauwen2007overview}. Furthermore, classical or quantum physical systems are also amenable to be used as the dynamical system\cite{tanaka2019recent,mujal2021opportunities,nakajima2021reservoir}, paving the way to harvesting computational power from essentially random physical systems with fading memory and complex dynamics. In RC such systems are usually called reservoirs.

After recent seminal works investigating the suitability of the transverse-field Ising model for RC purposes \cite{fujii2017harnessing,chen2019learning}, there has been a surge of interest in the quantum case in particular. Indeed, the initial model has been refined and analyzed in several ways\cite{nakajima2019boosting,Kutvonen2020,martinez2020information,martinez2021}, whereas new proposals have introduced RC based on quantum circuits\cite{9029180,PhysRevApplied.14.024065}, NMR systems\cite{negoro2018machine} and continuous variable quantum systems\cite{nokkala2021gaussian}. The results have been promising, suggesting that both in the discrete and continuous variable case quantum reservoirs may have an advantage over their classical counterparts in terms of how rapidly the potential reservoir performance improves with size\cite{fujii2017harnessing,martinez2020information,nokkala2021gaussian}. One of the biggest hurdles is in fact the extraction of the classical output from the quantum systems, since not only does a single measurement reveal only a tiny amount of information about a quantum system in an unknown state, it also alters the state and therefore competes with the inputs in driving reservoir dynamics. For certain special systems, such as NMR systems, an enormous amount of copies of the reservoir are naturally available which has been proposed to allow one to bypass the measurement back-action problem when collective input injections and measurements can be carried out\cite{negoro2018machine}. In general, repeatedly initializing and subsequently measuring a quantum system extracts classical information out of it---for example, a value of an observable---however such an approach is far from ideal for time series processing for two reasons. Firstly, carrying out the repetitions anew for every element in the output time series one wishes to learn introduces severe overhead, and secondly, it is hard to imagine how such a protocol can run in an online mode, continuously producing elements of the output time series. All in all, output extraction is a major challenge in exploiting the potential of quantum reservoirs for classical time series processing. 

Here we lay down an alternative, RC inspired path that can fully harness the quantumness of the reservoir while largely sidestepping the measurement problem. Namely, we introduce online time series processing with random fading memory quantum systems where both the input and desired output time series consist of quantum information. Such temporal quantum tasks are by construction impossible for a classical reservoir, whereas no measurements are required after the reservoir has been trained since the output can remain quantum. Specifically, we consider random networks of interacting quantum harmonic oscillators as the reservoir. Taking inspiration from RC, we train only the interaction Hamiltonian between the network and the carriers of input information. The main difference with RC is how the output is formed; in the former it is a trained function of reservoir observables, here the output is imprinted directly on the quantum systems acting as carriers of data. This affects also the training process as will be seen.

We illustrate the possibilities of the proposed model with three different temporal tasks. The short term quantum memory (STQM) task is the quantum analog of the short term memory task\cite{jaeger2002short} commonly used as a benchmark task in classical RC---the objective is to recall past inputs that are no longer available using the memory of the reservoir. Another common task is channel equalization task\cite{mathews1994adaptive}, where the input time series is transmitted through a noisy nonlinear channel that also mixes the time series with various echoes of itself, and the objective is to recover the original time series from the distorted one. Here we generalize it to inverting the transformation caused by a quantum channel. Despite being generalizations of a classical task it will be seen that the quantum cases have notable differences. Furthermore, we introduce a task without a classical counterpart which we call the entangler. Here the objective is to create entanglement between different initially uncorrelated systems by letting each of them in turn interact once with the reservoir but never between each other. We find that all these tasks are possible to solve using random untrained networks of interacting quantum harmonic oscillators; remarkably, not even the network initial state needs to be controlled. Finally, we briefly discuss partial generalizations, i.e. cases where only the input or only the output time series is quantum.

\section*{Results}


\subsection*{The model}

In RC a reservoir is a dynamical system that can be steered by an input time series to a trajectory in its state space determined by the inputs alone, i.e. its internal variables become completely determined by the input history at the limit of many inputs. If the variables can be monitored then the response of the reservoir at different timesteps can be post-processed to achieve a desired transformation from the input time series to an output time series. Importantly, for sufficiently complex reservoirs nontrivial transformations can be achieved by cheap post-processing, such as a linear combination of the variables.

Here the reservoir is a network of $N$ unit mass quantum harmonic oscillators interacting with springlike couplings. Such units are used that $\hbar=1$ and $k_B=1$. Let $\mathbf{p}^\top=\{p_1,p_2,\ldots,p_N\}$ and $\mathbf{q}^\top=\{q_1,q_2,\ldots,q_N\}$ be the vectors of momentum and position operators of the oscillators. The reservoir Hamiltonian $H_R$ is
\begin{equation}
H_R=\dfrac{\mathbf{p}^\top\mathbf{p}}{2}+\dfrac{\mathbf{q}^\top(\boldsymbol{\Delta}_{\boldsymbol{\omega}}^2+\mathbf{L})\mathbf{q}}{2},
\label{eq:reservoir}
\end{equation}
where the diagonal matrix $\boldsymbol{\Delta}_{\boldsymbol{\omega}}$ holds the oscillator frequencies $\boldsymbol{\omega}^\top=\{\omega_1,\omega_2,\ldots,\omega_N\}$ and the symmetric matrix $\mathbf{L}$ has elements $\mathbf{L}_{ij}=\delta_{ij}\sum_k g_{ik}-(1-\delta_{ij})g_{ij}$. Here $g_{ij}\geq 0$ are interaction strengths between the reservoir oscillators. Aside from oscillator frequencies there is a one-to-one correspondence between $H_R$ and weighted simple graphs. Indeed, $\mathbf{L}$ can be interpreted as the Laplace matrix of such a graph, and a given graph with a Laplace matrix $\mathbf{L}$ defines $H_R$ through Eq.~\eqref{eq:reservoir}. 

We consider temporal quantum tasks (analogous to temporal tasks in RC), which we define in this work as follows. The input time series $\mathbf{s}=\{\ldots,\rho_{m-1}^{I},\rho_{m}^{I},\rho_{m+1}^{I},\ldots\}$ consists of quantum states $\rho_{m}^{I}$ where $m$ indicates the timestep. In general these can be states of multimode continuous variable quantum systems, however we assume that apart from their states the systems are identical, i.e. each system has the same Hamiltonian $H_S$. The input time series is processed by the reservoir into output time series $\mathbf{o}=\{\ldots,\rho_{m-1}^{O},\rho_{m}^{O},\rho_{m+1}^{O},\ldots\}$ by letting each system in turn interact with the reservoir for some time $\Delta t$ according to an interaction Hamiltonian $H_I$ coupling every reservoir oscillator to every subsystem. The order of interactions is given by the timesteps. Consequently $\mathbf{o}$ is the image of $\mathbf{s}$ and reservoir initial conditions under a transformation induced by the full Hamiltonian $H=H_R+H_S+H_I$ and the interaction time $\Delta t$. In a temporal quantum task we attempt to realize a given transformation from $\mathbf{s}$ to $\mathbf{o}$ in this way. Besides the uncorrelated case one may also consider correlations between the systems at different timesteps, and we will return to this point later. 

In the special case where $H_S$ consist of $M$ unit mass quantum harmonic oscillators and the interactions in $H_I$ are springlike couplings, $H$ has the same general form as $H_R$. The transformation induced by $H$ and $\Delta t$ on the operators of the reservoir and input system is now linear and can be given in terms of a symplectic matrix $\mathbf{S}$. Let $\mathbf{x}^R_k$ be the form of the reservoir operators after $k$-th input has been processed, let $\mathbf{x}^S_k$ be the operators of the $k$-th input and let $\mathbf{x}^O_k$ be the operators of the $k$-th output. Now
\begin{equation}
\begin{pmatrix}
\mathbf{x}^R_{k+1} \\
\mathbf{x}^O_{k+1}
\end{pmatrix}=
\mathbf{S}
\begin{pmatrix}
\mathbf{x}^R_{k} \\
\mathbf{x}^S_{k+1}
\end{pmatrix}=
\begin{pmatrix}
\mathbf{A} & \mathbf{B} \\
\mathbf{C} & \mathbf{D} 
\end{pmatrix}
\begin{pmatrix}
\mathbf{x}^R_k \\
\mathbf{x}^S_{k+1}
\label{eq:sympmatstep}
\end{pmatrix},
\end{equation}
where the symplectic matrix has been divided into blocks such that $\mathbf{A}$ is $2N\times 2N$ and $\mathbf{D}$ is $2M\times 2M$. By iterating this equation we immediately get the form of both the reservoir and input modes for some timestep $m$:
\begin{equation}
\begin{cases}
\mathbf{x}^R_m=\mathbf{A}^m\mathbf{x}^R_0+\sum_{k=1}^m\mathbf{A}^{m-k}\mathbf{B}\mathbf{x}^I_k,\\
\mathbf{x}^O_m=\mathbf{C}\mathbf{x}^R_{m-1}+\mathbf{D}\mathbf{x}^I_{m},
\end{cases}
\label{eq:fullsystemdynamics}
\end{equation}
where $\mathbf{x}_0^R$ is the initial form of the reservoir modes. The form of the output modes for some timestep $m$ as a function of $\mathbf{x}_0^R$ and input history is then
\begin{equation}
\begin{split}
\mathbf{x}^O_m &=\mathbf{C}\mathbf{A}^{m-1}\mathbf{x}^R_0+\mathbf{D}\mathbf{x}^I_{m}+\mathbf{C}\sum_{k=1}^{m-1}\mathbf{A}^{m-k-1}\mathbf{B}\mathbf{x}^I_k \\
&\approx \mathbf{D}\mathbf{x}^I_{m}+\mathbf{C}\sum_{k=1}^{m-1}\mathbf{A}^{m-k-1}\mathbf{B}\mathbf{x}^I_k \quad\text{when $\rho(\mathbf{A})<1$ and $m\gg 1$},
\end{split}
\label{eq:QQinputdynamics}
\end{equation}
where the first line is exact and the second line an approximation which holds when the spectral radius $\rho(\mathbf{A})$---not to be confused with quantum states---is less than 1 and enough inputs have been processed. Given that the equations of motion are conveniently expressed in terms of the operators, in the rest of this manuscript we will simply write $\mathbf{s}=\{\ldots,\mathbf{x}_{m-1}^{I},\mathbf{x}_{m}^{I},\mathbf{x}_{m+1}^{I},\ldots\}$ and $\mathbf{o}=\{\ldots,\mathbf{x}_{m-1}^{O},\mathbf{x}_{m}^{O},\mathbf{x}_{m+1}^{O},\ldots\}$.

When $\rho(\mathbf{A})<1$ the output time series $\mathbf{o}$ becomes independent of the initial conditions at the limit of infinitely long input history, which is known as the echo state property\cite{jaeger2001echo} in the RC literature. In fact, it can be shown\cite{nokkala2021gaussian} that satisfying the spectral radius condition gives the reservoir also the so-called fading memory property\cite{boyd1985fading}, which guarantees that $\mathbf{x}^R_m$ and therefore $\mathbf{x}^O_m$ become well-approximated by a continuous function of only a finite number of past inputs at the limit $m\gg 1$. This not only ensures that the initial conditions can be ignored but also prevents any physical quantities from diverging: the reservoir state never leaves the state space as long as all input states are physical, not even in the limit $m\to\infty$. At variance, if $\rho(\mathbf{A})\geq1$ then, e.g., reservoir excitations may diverge. Finally, in the special case where $\mathbf{A}$ is nilpotent for some index $n$ there is sudden death of reservoir memory where $\mathbf{x}^R_m$ is a function of exactly $n$ previous inputs. As the index of a nilpotent matrix is always at most its order\cite{zhang2011matrix}, $n\leq 2N$. 

\subsection*{RC inspired quantum time series processing}

The system given in Eqs.~\eqref{eq:fullsystemdynamics} can be harnessed for RC by considering only $\mathbf{x}^R_m$ when it is assumed that the states in $\mathbf{s}$ are in fact functions of the elements of a classical time series\cite{nokkala2021gaussian}. When $\rho(\mathbf{A})<1$ there is fading memory and the observables of the reservoir become well-approximated by continuous functions of only a finite number of past inputs. Different transformations to a classical output time series can then be accomplished by training a simple function of the reservoir observables such as first moments, second moments or covariances of $\mathbf{x}^R_m$. The resulting RC can be analyzed with contemporary RC theory since the latter is agnostic to the mechanism that creates the functions of the input. Importantly, $H_R$ is not trained and can be random since $\rho(\mathbf{A})<1$ can typically be achieved by just tuning $\Delta t$.

\begin{figure}[ht]
\centering
\includegraphics[trim=0.5cm 5cm 5.15cm 0.5cm,clip=true,width=0.95\linewidth]{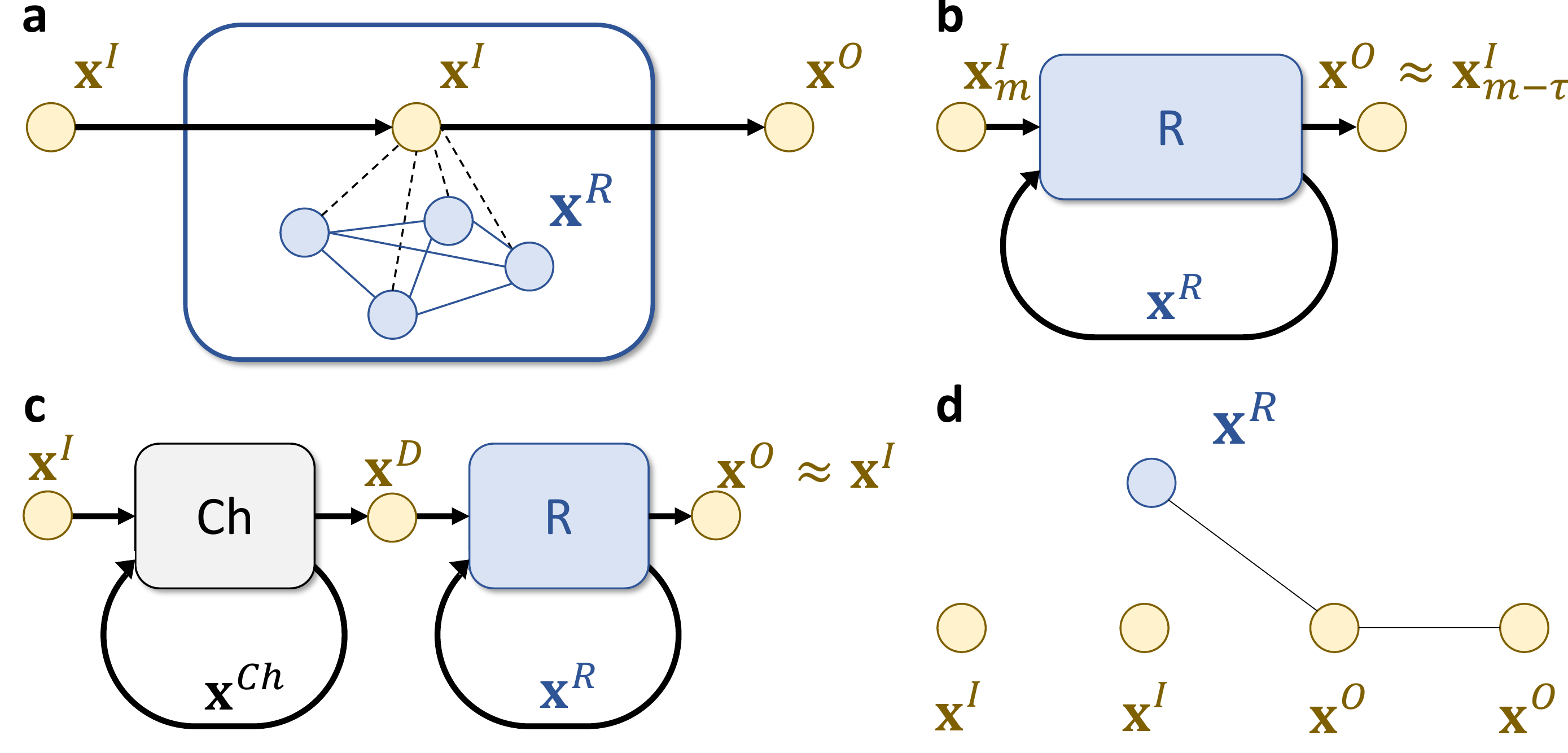}
\caption{\textbf{a} A time series of quantum systems with operators $\mathbf{x}^I$ is processed into another time series where the transformed operators are indicated by $\mathbf{x}^O$. Each system interacts one after another with a random network of oscillators---later called the reservoir---with operators $\mathbf{x}^R$. In general the reservoir state depends on the states of systems it has interacted with, in turn making $\mathbf{x}^O$ a function of all previous $\mathbf{x}^I$. Different transformations can be achieved by tuning only the interaction terms, indicated by dashed black lines. \textbf{b} In the short term quantum memory task the objective is to transform $\mathbf{x}^I$ at timestep $m$ into $\mathbf{x}^I$ at timestep $m-\tau$ where $\tau\geq 0$ is a delay. This is possible when the sought state can be distilled out of the reservoir memory, stored in $\mathbf{x}^R$. \textbf{c} In the quantum channel equalization task the input is transformed by a random, uncontrollable system with operators $\mathbf{x}^{Ch}$ into the distorted input $\mathbf{x}^D$. The reservoir is to recover the original input from the distorted one. \textbf{d} In the entangler task the reservoir is to entangle the systems in the time series. Like before the systems never directly interact and only one of them interacts with the reservoir at any given time. Here there is only one reservoir oscillator and only a part of the input and output time series is shown for simplicity, whereas entanglement is indicated by solid black lines.}
\label{fig:schematic}
\end{figure}

The scheme for using the reservoir to process temporal quantum information instead is shown in Fig.~\ref{fig:schematic}\textbf{a}. Here $\mathbf{s}$ itself will be the input while $\mathbf{o}$ consists of quantum information encoded in $\mathbf{x}^O_m$. Thanks to fading memory, each $\mathbf{x}^O_m$ is completely determined by $\mathbf{x}^I_i$ where $i\leq m$ and the initial reservoir state can be ignored. To achieve different transformations $\mathbf{s}\mapsto\mathbf{o}$, the matrix $\mathbf{S}$ induced by $H=H_R+H_S+H_I$ and $\Delta t$ must be changed while preserving $\rho(\mathbf{A})<1$. Whereas the number of parameters in $H$ is proportional to $(N+M)^2$, we take an approach inspired by RC and attempt to achieve online quantum time series processing by only training the $NM$ interaction terms in $H_I$, leaving both $H_R$ and $H_S$ fixed.  We will provide strong numerical evidence that this is in practice enough to succeed in many different temporal quantum tasks, even with random $H_R$. 

The similarity of this result with the fact that in RC with fading memory systems it is enough to only train the final weights in the readout layer is uncanny, since the situations are quite different. Indeed, unlike in RC here reservoir dynamics cannot be separated from training since tuning $H_I$ changes the dynamics. We are in fact not aware of any theoretical results that could be used to explain the phenomenon. In the following we provide a brief overview of the general purpose training process. See Methods for further details.

First the input time series $\mathbf{s}$ is divided into three phases: the preparation, training and test phases. The role of the preparation phase is to get rid of the influence of the reservoir initial state. During the training phase the performance is checked using a cost or objective function, which varies depending on the task but for a fixed input and reservoir is completely determined by $H_I$. The specific forms will be introduced along with the respective tasks. The interaction Hamiltonian is varied to optimize the function using a simple stochastic function optimizer. Spectral radius condition is enforced by providing the optimizer as initial points only such $H_I$ that the condition is satisfied; during minimization points that violate the condition can be expected to perform worse and are therefore discarded. Unless the time between inputs $\Delta t$ is fixed by the task, training may be repeated for different choices of $\Delta t$. In the test phase trained $H_I$ and best $\Delta t$ are used and reservoir output is collected to check the performance using a task dependent figure of merit. In this way the trained reservoir is exposed to new input and must be able to generalize beyond the specific inputs in the training phase to succeed. These are the results shown in the figures.

\subsection*{Examples of temporal quantum tasks}

\subsubsection*{Parameter values used in numerical experiments}

In all numerical experiments reported throughout the section we consider as the reservoir random completely connected networks of $N$ identical oscillators with a bare frequency $\omega_0=0.25$. Each coupling strength between the reservoir oscillators is $g_{ij}\in[0,0.2]$, chosen uniformly at random. The $M$ input modes are are also such oscillators.  Although in principle having $\rho(\mathbf{A})<1$ is enough for fading memory, in practice we impose the limit $\rho(\mathbf{A})<0.99$ to avoid issues with finite precision numerics. The influence of the initial state of the reservoir is washed out during the preparation phase and is therefore irrelevant, however in practice we use the ground state of $H_R$. For each different case we show results of 100 random realizations of the reservoir and when applicable, other quantities such as the input time series $\mathbf{s}$. In all cases the lengths of preparation, training and test phases are 40, 80 and 40, respectively. The values of the time between inputs $\Delta t$ and the input states themselves vary and will be reported along the tasks.

\subsubsection*{Short term quantum memory task}

The short term memory task is a paradigmatic task in classical RC where the input $s_k$ at some timestep $k$ is a real scalar or vector, and the target is $s_{k-\tau}$ where $\tau$ is the delay. Checking the performance in this task for different delays is often used to gauge how much linear memory a reservoir has. The short term quantum memory (STQM) task is its direct generalization where $s_k$ is a state of a quantum system. Specifically,
\begin{equation}
 \begin{cases} 
 \mathbf{x}^I_k, & \text{(input at timestep $k$)} \\ 
\mathbf{x}^O_k\approx\mathbf{x}^I_{k-\tau}. & \text{(target at timestep $k$)} 
 \end{cases}
 \label{eq:STQMtask}
\end{equation}
The objective is to achieve the transformation $\mathbf{s}\mapsto\mathbf{o}$ defined by Eq.~\eqref{eq:STQMtask} by training $\Delta t$ and Hamiltonian $H_I$. $H_R$ is assumed to be random but fixed and $\mathbf{x}_0^R$ can be arbitrary. Unlike in the classical case where the amount of information grows linearly with input size, here the growth is more rapid as the reservoir must be able to delay also the correlations and entanglement between different input modes in $\mathbf{x}_k^I$. We mention in passing that the process tomography of such a delay map in the discrete variable case was considered in Ref.~[\citeonline{tran2021learning}].

It should be pointed out that for single-mode input states the task can in principle be done exactly for any delay $\tau$ by concatenating $N=\tau$ single oscillator reservoirs, which can be compared to deep RC. For $\tau=1$ the states of the reservoir and input must be swapped; the required interaction strength and time can be solved analytically. Clearly using a single reservoir oscillator with double the interaction time solves the task for $\tau=0$ since the states are swapped twice, while using a sequence of swaps with $N$ different single mode reservoirs achieves a delay $\tau=N$. For $M>1$ and $\tau>0$ the task can be done by using $N=M\tau$ non-interacting reservoir oscillators, provided that the input systems are likewise non-interacting. Here we show that the task can in fact be solved by using a single random and fixed reservoir and letting the input interact only once with it, which can be compared to ordinary (shallow) RC.

The simplicity of this task makes it amenable to a special purpose training procedure. Indeed, from Eq.~\eqref{eq:QQinputdynamics} the following conditions to solve the task can be observed:
\begin{equation}
\begin{cases}
\mathbf{D}\approx\mathbf{I},\ \mathbf{C}\mathbf{A}^t\mathbf{B}\approx\mathbf{0}\ \forall t\geq 0 & \textrm{if $\tau=0$,}\\
\mathbf{C}\mathbf{A}^{\tau-1}\mathbf{B}\approx\mathbf{I},\ \mathbf{D}\approx\mathbf{C}\mathbf{A}^{t\neq\tau-1}\mathbf{B}\approx\mathbf{0}& \textrm{if $\tau>0$.}
\end{cases}
\label{eq:STQMcondition}
\end{equation}
When satisfied the contributions from the incorrect timesteps are suppressed while the contribution from the correct one is enhanced. The conditions in Eqs.~\ref{eq:STQMcondition} for the full symplectic matrix induced by all three Hamiltonians must be satisfied by training only one of them, $H_I$. In practice the cost function given by Eqs.~\ref{eq:STQMobjective} in Methods is minimized to train the reservoir. Additionally, training of $H_I$ is repeated for $\Delta t= 2\pi/\omega_0,4\pi/\omega_0,\ldots,8\pi/\omega_0$ and the best value is chosen for testing phase.  Notably, training is input state independent; although the training phase in $\mathbf{s}$ could therefore be omitted, it is kept for the sake of consistency. It should be stressed that since the task is linear in the involved modes for any state, if training is successful the reservoir can delay also non-Gaussian $M$-mode states.

\begin{figure}[t]
\centering
\includegraphics[trim=0cm 0cm 0cm 0cm,clip=true,width=0.95\linewidth]{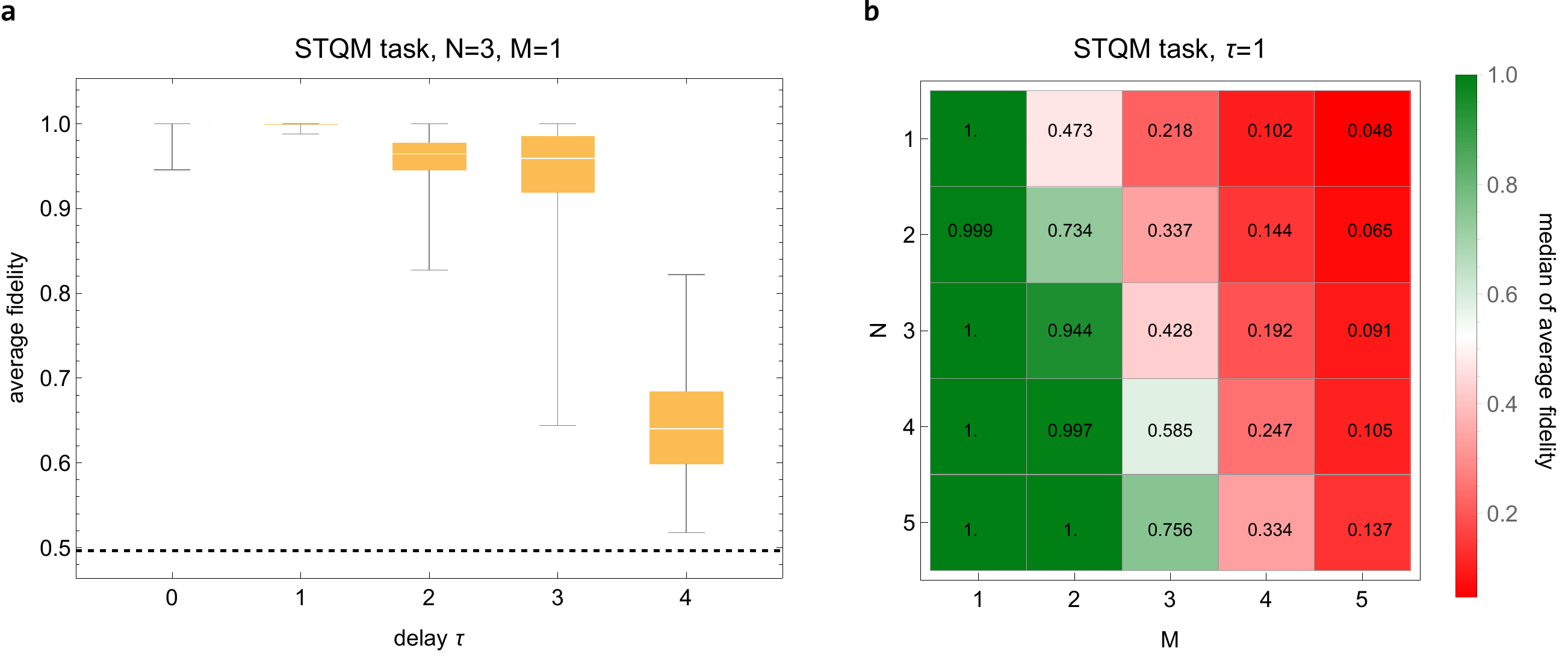}
\caption{Results for STQM task. Different reservoir sizes $N$, input sizes $M$ and delays $\tau$ are considered, and for each different set of values 100 random reservoirs were used. \textbf{a} Delay $\tau$ is varied for fixed $N$ and $M$ and all results are shown with a box plot. The box plot shows the minimum (lower whisker), maximum (upper whisker), median (line between boxes) and the first and third quartiles (beginning of lower box and end of the upper box, respectively). The fidelity achieved by random guessing is indicated by the dashed horizontal line. \textbf{b} Delay $\tau$ is fixed while both $N$ and $M$ are varied. Here only the median value is shown. See text for details.}
\label{fig:STQM}
\end{figure}

Results of numerical experiments are shown in Fig.~\ref{fig:STQM}. The input time series $\mathbf{s}$ consists of random zero mean $M$-mode Gaussian states generated by acting with a random symplectic matrix on a thermal state of non-interacting oscillators. For details and parameter values used see Methods; it should be stressed that in general this creates correlated states. The reservoir parameters are as specified previously. Different values of $N$, $M$ and delay $\tau$ are considered, whereas the figure of merit is the (Uhlmann) fidelity averaged over the test phase, possible to calculate in closed form for arbitrary Gaussian states applying, e.g., results of Ref.~[\citeonline{banchi2015quantum}].

In panel \textbf{a} reservoir size $N$ and input size $M$ have been fixed and the delay $\tau$ is varied. The performance is excellent especially for small delays $\tau=0,1,2$ where the median values are $\bar{F}>0.999999$, $\bar{F}>0.9999$ and $\bar{F}>0.964$, respectively. Even the worst possible performance is decent for small delays, and it is conceivable that if in their case training was repeated with a different seed for the random number generator the performance might be improved significantly. In panel \textbf{b} the delay is fixed and $N$ and $M$ are varied; the diagonal part of the array shows the case $N=M$. As expected, the reservoir struggles significantly with the task when $N<M$ whereas for $N>M$ the performance is mostly good. Remarkably, a random reservoir can be trained to delay also multimode states by only tuning $H_I$ and $\Delta t$. 

\subsubsection*{Quantum channel equalization}

This task is inspired by the channel equalization task where the input time series is distorted by a transmission through a classical channel with fading memory and the objective is to invert the transformation by the channel and restore the original time series. The quantum counterpart is presented in Fig.~\ref{fig:schematic}\textbf{c}, where the original time series consists of states of quantum systems with operators $\mathbf{x}^I$ that are transformed by an interaction with some fixed but random system with operators $\mathbf{x}^{Ch}$, which constitutes the channel. We assume the channel to induce a linear transformation of the modes given by some symplectic matrix, i.e. the relevant Hamiltonians are quadratic, and furthermore that it has fading memory, which implies that $\mathbf{x}^{Ch}$ is well-approximated by a function of a finite number of past inputs. The transformed operators of the input system are denoted by $\mathbf{x}^D$. A reservoir with operators $\mathbf{x}^R$ is trained to recover the original time series from $\mathbf{x}^D$. The equations of motion can be recast in the same general form as before as explained in Methods.  

It should be stressed that simply training the reservoir to perform the inverse of the channel symplectic matrix will not work since the channel acts on $\mathbf{x}^{Ch}$ and $\mathbf{x}^I$ but the reservoir symplectic matrix $\mathbf{S}$ acts on $\mathbf{x}^R$ and $\mathbf{x}^D$. The transformed input $\mathbf{x}^D_k$ for some timestep $k$ in general does not contain full information of any of the previous inputs because part of the information is in $\mathbf{x}^{Ch}$ and in general also in the correlations between the channel and the distorted input. Since it is well known that unknown quantum states can neither be cloned nor amplified the task as given is in fact impossible---in stark contrast with its classical counterpart. To have any hope of success, the task must be modified. 

Here we will do this using techniques inspired by classical RC. Instead of a single copy of the input state we will transmit a product state of $\mathfrak{s}$ copies, but still require the reservoir to distill only a single copy of the original input, thus giving the reservoir additional quantum information to work with. We call this spatial multiplexing at order $\mathfrak{s}$. Additionally, we transmit the same product state $\mathfrak{m}$ times, one copy after another, requiring the reservoir to output the original input only after the $\mathfrak{m}$-th copy. We call this temporal multiplexing at order $\mathfrak{m}$. More formally,
\begin{equation}
 \begin{cases}
 \bigoplus_{i=1}^{\mathfrak{s}}\mathbf{x}_k^I,\:\text{$\mathfrak{m}$ copies sequentially}& \text{(inputs to channel at timestep $k$)}\\
 \mathbf{x}^D_{k,1},\:\mathbf{x}^D_{k,2},\ldots\mathbf{x}^D_{k,\mathfrak{m}}\:\text{sequentially} & \text{(inputs to reservoir at timestep $k$)} \\ 
\mathbf{x}^O_k\approx\mathbf{x}^I_{k}, & \text{(target at timestep $k$)} 
 \end{cases}
 \label{eq:Qchantask}
\end{equation}
Here the timesteps are to be understood as the points where we switch from one set of identical copies to another, as determined by the original unaltered input time series $\mathbf{s}$. As a final remark before moving on, if $\mathbf{s}$ is given but unknown neither spatial nor temporal multiplexing can be used, making the task again impossible. It must be assumed that there is a source that directly generates states according to some $\mathfrak{s}$ and some $\mathfrak{m}$, or alternatively that the states are known to the sender who may then prepare the copies. It may be asked if some other modifications could help solve the task even for unknown $\mathbf{s}$, but this is outside the scope of present work.

\begin{figure}[t]
\centering
\includegraphics[trim=0cm 0cm 0cm 0cm,clip=true,width=0.95\linewidth]{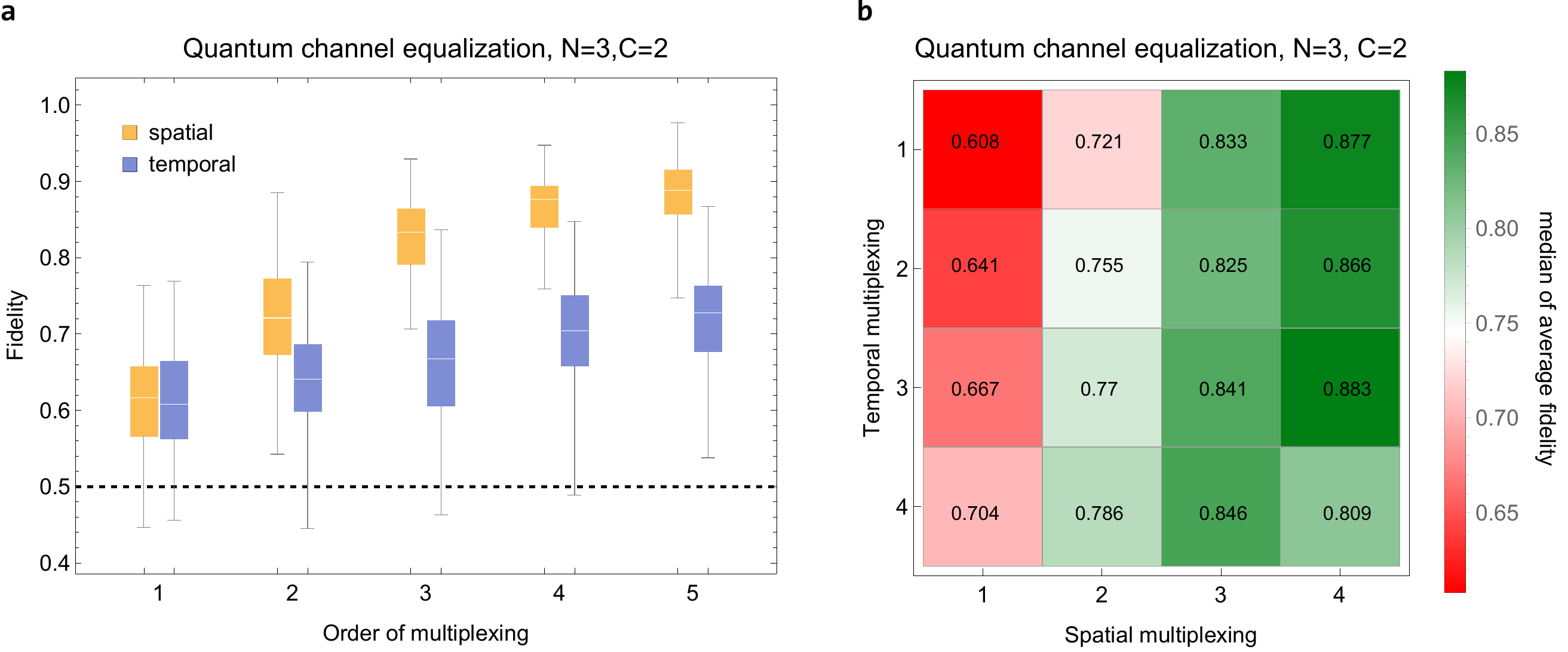}
\caption{Results for the quantum channel equalization task. Reservoir, input and channel sizes are fixed to $N=3$, $M=1$ and $C=2$, respectively, while the orders of spatial and temporal multiplexing of the input are varied. In the former product states of identical copies of the input are used instead of single mode states, and in the latter, identical copies of the input are injected sequentially. \textbf{a} The two multiplexings are used separately with results shown with a box plot as in Fig.~\ref{fig:STQM}. The fidelity achieved by random guessing is indicated by the dashed horizontal line. \textbf{b} The two multiplexings are used together. In all cases results of 100 random realizations of the reservoir, input and channel have been used.}
\label{fig:Qchan}
\end{figure}

With the analysis complete for now, we move on to numerical experiments. The input consists of single mode zero mean Gaussian states. The channel has $C=2$ oscillators and the Hamiltonian has the same general form as the reservoir Hamiltonian. This system is taken to interact with the inputs such that the spectral radius of the relevant block in the symplectic matrix is at most $0.95$. For the reservoir $N=3$. The input time series consists of random zero mean Gaussian states like before in the STQM task, however for simplicity we focus only on the single mode case where $M=1$. Unlike before, the time between inputs is taken to be fixed by the channel and is therefore kept at a constant value which was chosen to be $\Delta t=1.5\pi/\omega_0$. Results are shown in Fig.~\ref{fig:Qchan}.

In panel \textbf{a} spatial and temporal multiplexing are considered separately. At $\mathfrak{s}=\mathfrak{m}=1$ both reduce to the original formulation of the task, which was already concluded to be unsolvable. Still, performance exceeds that of random guessing. Performance increases quickly with $\mathfrak{s}$ and slowly with $\mathfrak{m}$. Indeed, already at $\mathfrak{s}=2$ results tend to be better than at $\mathfrak{m}=5$. That being said, unlike increasing $\mathfrak{m}$ increasing $\mathfrak{s}$ increases the number of terms in $H_I$ which, e.g., makes training slower. In panel \textbf{b} spatial and temporal multiplexing are considered together. Curiously, performance does not always increase with $\mathfrak{m}$ and $\mathfrak{s}$. For example, both $\mathfrak{m}=3,\;\mathfrak{s}=4$ and $\mathfrak{m}=4,\;\mathfrak{s}=3$ lead to a better performance than $\mathfrak{m}=4,\;\mathfrak{s}=4$. Moreover, spatial multiplexing alone achieves a performance not too far off from the best case.

\subsubsection*{Entangler}

\begin{figure}[ht]
\centering
\includegraphics[trim=0cm 0cm 0cm 0cm,clip=true,width=0.95\linewidth]{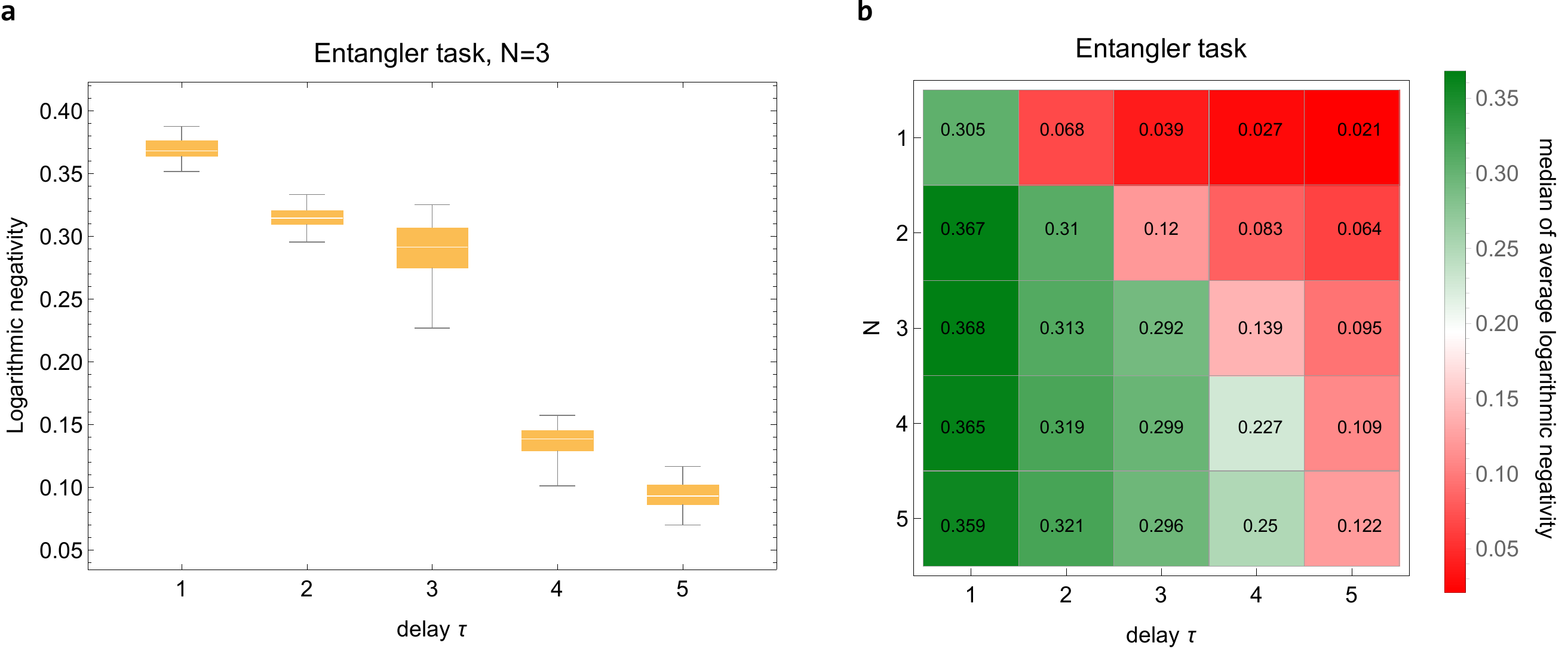}
\caption{Results for entangler task. For each different case 100 random networks were created, whereas the figure of merit is average logarithmic negativity. \textbf{a} Delay is varied for fixed $N$ and all results are shown with a box plot as in Fig.~\ref{fig:STQM}. Delay $1$ corresponds to nearest neighbors, $2$ to next nearest neighbors and so on. \textbf{b} Median logarithmic negativity when $N$ and delay are varied. See text for details.}
\label{fig:entangler}
\end{figure}

In this task the inputs $\mathbf{s}_k$ are taken to be uncorrelated single mode systems in the vacuum state whereas the target time series consists of entangled states. Although more complicated patterns can be envisioned, here we focus on entangling systems with a fixed delay $\tau$. That is to say the goal is to entangle $\mathbf{s}_k$ with $\mathbf{s}_{k-\tau}$ by training $H_I$. For $\tau=1$ the target is then a chain of systems with nearest neighbor connections where the connections are entanglement, for $\tau=2$ a chain with next nearest neighbors connected and so on. Importantly, we assume that only one input interacts with the reservoir at any given timestep and all the others are unavailable, and furthermore assume that there are never any direct interactions between the systems in $\mathbf{s}$. In fact, if we imagine that the systems are periodically generated by a source of vacuum states then the input $\mathbf{s}_{k+1}$ does not even exist at some timestep $k$. A system or a device that can solve the task can turn the source of uncorrelated states into one of entangled states.

Much like in the STQM task, in certain special cases and allowing for time-dependent $H_I$ the task can be achieved analytically and exactly, as depicted in Fig.~\ref{fig:schematic}\textbf{d} where only the case $\tau=1$ is considered for simplicity. At each timestep the reservoir and the ancilla are first entangled using an interaction Hamiltonian of one form and then their states are swapped using a different interaction Hamiltonian; in both cases one can analytically solve what the precise form of the interactions and interaction times must be. In fact, since the operations commute the order does not matter. Every application of the entangling gate creates a link in Fig.~\ref{fig:schematic}\textbf{d}, which is later swapped to the next input system. The role of the reservoir oscillator is to provide short term quantum memory. Without it, the task becomes impossible.

Here we solve the task with the RC inspired approach by training the time-independent $H_I$ to maximize the entanglement, as quantified by logarithmic negativity\cite{vidal2002computable} between $\mathbf{s}_k$ and $\mathbf{s}_{k-\tau}$. Like in the STQM task, training is repeated for $\Delta t= 2\pi/\omega_0,4\pi/\omega_0,\ldots,8\pi/\omega_0$ and the best value is chosen for testing phase. To succeed the reservoir must simultaneously create entanglement and re-distribute the quantum information correctly as explained previously. Results are shown in Fig.~\ref{fig:entangler}. In panel \textbf{a} the reservoir size is fixed to $N=3$ and the delay $\tau$ is varied, and the logarithmic negativity averaged over the systems in the test phase is shown. The logarithmic negativity achieved for the shortest delay is between $0.35$ and $0.4$, which corresponds to that of a twin beam state with two-mode squeezing parameter $0.175\leq s\leq 0.2$. Performance decreases slowly up to $N=\tau$, but then collapses for delays $\tau>N$, bearing a striking similarity with panel \textbf{a} of Fig.~\ref{fig:STQM}. One may interpret this as the reservoir being able to remember up to $N$ single mode states before running out of memory. The same behaviour can be observed also in panel \textbf{b} where both $N$ and delay $\tau$ are varied and median performances are shown.

\subsection*{Partial generalizations}

\begin{figure}[t]
\centering
\includegraphics[trim=0cm 0cm 0cm 0cm,clip=true,width=0.95\linewidth]{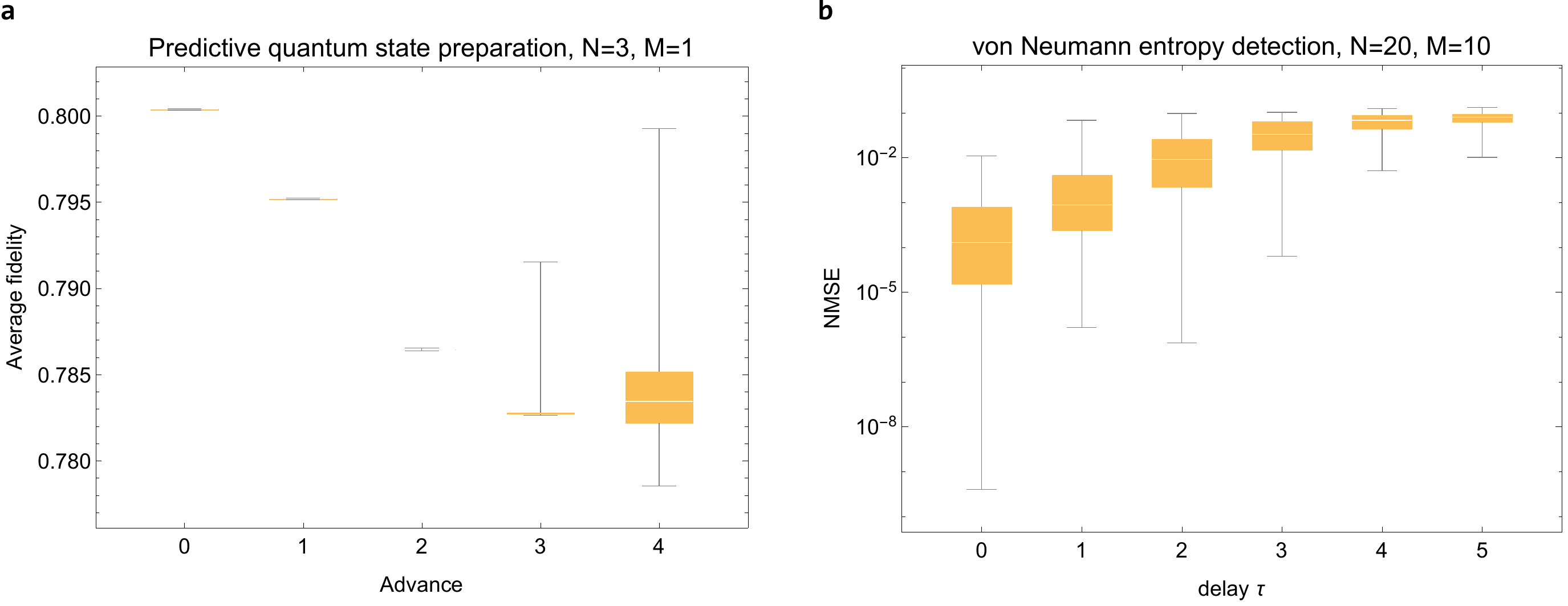}
\caption{Partial generalizations. \textbf{a} Results for predictive quantum state preparation where the input is classical but target output is quantum. The reservoir is trained to prepare the quantum state according to future classical inputs which it must deduce from previous inputs. Specifically, the inputs are thermal states where the number of thermal excitations follows the well known Santa Fe time series, and the targets are squeezed vacuum states where the squeezing parameter is to coincide with the number of thermal excitations of a future input.
\textbf{b} Results for von Neumann entropy detection where the input is quantum but the target output is classical. Here the reservoir is trained to estimate the determinants of input covariance matrices---which completely determines, e.g., the von Neumann entropy---for different delays. Random product states of $M=10$ identical single mode states are used as input. If reservoir observables are available for a single timestep, then the von Neumann entropies of multiple previous inputs can be estimated with a very low error.}
\label{fig:QCandCQ}
\end{figure}

In all previously introduced tasks both the input and the output time series consist of quantum information, however one may consider partial generalizations where one of them is still classical. Here we briefly illustrate the possibilities with two simple examples.

If the output is quantum but $\mathbf{s}$ is classical, say, a time series of systems in thermal states, one may follow the framework used previously. As an example task we consider predictive quantum state preparation where the reservoir is trained to prepare a given quantum state---here, squeezed vacuum---based on future classical inputs. For arbitrary $\mathbf{s}$ this is of course impossible, but if $\mathbf{s}$ is at least approximately predictable then the task can be in principle solved. Here we consider the Santa Fe chaotic time series, a dataset recorded from a far-infrared laser in a chaotic state\cite{huebner1989dimensions,weigend1993results} often used to benchmark the predictive power of classical reservoirs. Specifically, we normalize the Santa Fe time series and consider as $\mathbf{s}$ single mode thermal states such that the number of thermal excitations $(n_{\mathrm{th}})_k$ follows the normalized time series and the target is a squeezed vacuum state with a squeezing parameter $r_k=(n_{\mathrm{th}})_{k+a}$ where $a\geq 0$ is the advance, or the number of timesteps in the future the reservoir must be able to predict.

Results are shown in Fig.~\ref{fig:QCandCQ}\textbf{a}, where the average fidelity between the target state and the actual output state is shown for different values of the advance. There is very little spread for most values since only the reservoir is randomized between different realizations. Interestingly, there is an abrupt change in behavior when the advance $a$ exceeds the number of reservoir oscillators $N$. Even then the fidelity remains decent but there is considerably more spread in the performance.

In the opposite case where output is classical information, say, about the properties of the states carried by the input systems in $\mathbf{s}$, the approach where the classical output is formed from reservoir observables can be used, as outlined previously in the Model section. Let $\sigma(\mathbf{x}_k^R)$ be the reservoir covariance matrix at some timestep $k$. It can be shown\cite{nokkala2021gaussian} that
\begin{equation}
\begin{split}
    \sigma(\mathbf{x}_m^R)&=\mathbf{A}^m\sigma( \mathbf{x}_0^R)(\mathbf{A}^\top)^m+\sum_{k=1}^m\mathbf{A}^{m-k}\mathbf{B}\sigma( \mathbf{x}_k^I)\mathbf{B}^\top(\mathbf{A}^\top)^{m-k}\\&\approx\sum_{k=1}^m\mathbf{A}^{m-k}\mathbf{B}\sigma( \mathbf{x}_k^I)\mathbf{B}^\top(\mathbf{A}^\top)^{m-k}\quad\text{when $\rho(\mathbf{A})<1$ and $m\gg 1$},
\end{split}
\label{eq:covariances}
\end{equation}
which holds for any number $M$ of input modes. In principle, the elements of $\sigma(\mathbf{x}_k^R)$ can be estimated by performing measurements on multiple copies of the reservoir that has processed identical inputs $\mathbf{s}$, which however introduces substantial overhead. Analyzing just how much overhead is incurred is beyond the scope of this work; in what follows, it is assumed that the exact values of the elements of $\sigma(\mathbf{x}_k^R)$ are available. That being said, since the target is some function of reservoir observables the full Hamiltonian can remain constant, decoupling training from the dynamics. Indeed, once the elements are available multiple trained functions can be used to estimate a number of different features of $\mathbf{s}$.

As an example task, we consider as input random single mode Gaussian states and as target $\det(\sigma(\mathbf{x}^I_{k-\tau}))$, or the determinant of the single mode covariance matrix for some delay $\tau\geq 0$. This is an important quantity that, e.g., completely determines the purity, amount of thermal excitations and the von Neumann entropy of the state. Below we given an overview of the conditions under which this task was simulated; for full details, see Methods.

We consider a reservoir of size $N=20$ and consider $M=10$ input modes such that the input is in a random product state of identical single mode states. Furthermore, the Hamiltonian is such that only two reservoir oscillators interact with a single input mode, and there are no interactions between these triplets of two reservoir oscillators and a single input oscillator. The interaction strengths are random but fixed and the spectral radius condition is satisfied by tuning $\Delta t$. One may observe from Eq.~\eqref{eq:covariances} that $\sigma(\mathbf{x}_k^R)$ is linear in $\sigma(\mathbf{x}^I_{k-\tau})$ for any delay $\tau$ unlike the determinant, however this problem can be overcome by considering trained linear combinations of products of pairs of elements of $\sigma(\mathbf{x}_k^R)$. Here  the output is a trained linear combination of products of distinct pairs of the first row of $\sigma(\mathbf{x}_k^R)$, with training carried out as in Ref.~[\citeonline{nokkala2021gaussian}]. Finally, unlike elsewhere, we consider preparation, training and test phases of length $500$, $2000$ and $500$, respectively.

Results are shown in Fig.~\ref{fig:QCandCQ}\textbf{b}, where the the normalized mean squared error (NMSE) between the actual von Neumann entropy to that computed from the determinants estimated by the reservoir is shown. As can be seen, the NMSE is very small for all considered delays, suggesting an excellent agreement between the actual and predicted value. 

\section*{Discussion}

In this work we have introduced a RC inspired model for online processing of time series consisting of quantum information. Importantly, we have found that just with a judicious choice of the interaction Hamiltonian random instances of the model starting from any initial state can solve a variety of different tasks with high performance. We have also briefly illustrated the possibilities of partial generalizations of classical temporal tasks to cases where either the input or the target time series remains classical. Looking at the bigger picture, it is interesting to compare and contrast two distinct situations: when the output time series is to be quantum, and when it is to be classical. 

In the former case the output extraction problem hindering previous related work vanishes but engineering freedom is preserved: control of only a small subset of all parameters is sufficient for high performance. Furthermore, if the reservoir Hamiltonian is random but known then measurements are not needed even in the training stage provided one can simulate the dynamics. For the considered model in particular an unknown Hamiltonian can be probed first\cite{nokkala2016complex}. It can be imagined that a classical RC augmented with a state preparation mechanism could emulate the case where input is classical, but otherwise there is genuine quantumness: in the single shot case it is clear that no classical RC can emulate its quantum counterpart since the input data would first have to be transformed to classical information. That being said, training the interaction Hamiltonian is in general somewhat costly even when the dynamics can be simulated. If simulation is not possible the cost or objective function must be estimated with measurements, which should be expected to be a very challenging optimization problem in its own right by comparing with, e.g., variational quantum algorithms\cite{mcclean2014exploiting,wecker2015progress,babbush2018low,cai2020resource,garcia2021learning}. Finally, another advantage of RC is lost in multitasking where the same reservoir processing the same input can simultaneously solve many different tasks by using differently trained readout functions. In the quantum case any attempts to multitask will inevitably affect performance because there is only so much quantum information for forming the output.

When instead the output is classical, multitasking is possible and the training cost is minimal. The challenges are two-fold: the output extraction problem and pinpointing what role exactly quantumness plays aside from providing a larger state space. Moreover, even if the output extraction problem can be solved, the specific way it is solved may dictate what quantum systems are ultimately suitable. As we demonstrate here with von Neumann entropy detection, the case where input is quantum might be of particular interest however, since thanks to the memory and multitasking a plethora of information concerning multiple past input states can be distilled even if the reservoir observables are known only for one or few timesteps. This may be compared to recent proposals where information of only a single quantum state is extracted with the help of a larger quantum system and supervised machine learning\cite{ghosh2019quantum,ghosh2020reconstructing,angelatos2020reservoir}.

Indeed, the model proposed here and the results have created a fertile ground for further work in the direction where at least one of the time series is quantum. To the best of our knowledge there is currently little work on such temporal tasks, however the inverse problem of performing tomography of an unknown temporal quantum map has been recently considered in spin system\cite{tran2021learning}. Comparisons may also be made with a recent proposal to train quantum system to induce quantum gates between qubits\cite{ghosh2021realising}; its temporal generalization might consider gates between inputs at different timesteps, for example. Indeed, temporal quantum tasks could be tackled also in the discrete variable case. There is also room for further improvements for the introduced model by considering for example how training also the time between inputs can affect the performance or considering the case of non-Gaussian states or operations, which can be expected to lead to nonlinear memory\cite{braunstein2005quantum} where $\mathbf{x}^O_m$ can be nonlinear in $\mathbf{x}^I_i$ for $i\leq m$. One may also consider the prospects of a proof-of-principle experimental implementation since the general form of the reservoir Hamiltonian can in principle be realized in a multimode optics platform\cite{nokkala2018reconfigurable}.

\section*{Methods}

\subsection*{Generation of random zero mean Gaussian states}

In the single mode case where $M=1$ the states may be parameterized in terms of the thermal excitations $n_{\mathrm{th}}$, magnitude of squeezing $r$ and phase of squeezing $\varphi$. For displacement we consistently use $\alpha=0$, leading to the input first moments to vanish---the state is now completely characterized by its covariance matrix, which reads
\begin{equation}
\sigma(\mathbf{x}^I)=\frac{2n_{\textrm{th}}+1}{2}\begin{pmatrix}
(\cosh{(2r)}+\cos{(\varphi)}\sinh{(2r)})/\omega & \sin{(\varphi)}\sinh{(2r)} \\
\sin{(\varphi)}\sinh{(2r)} & (\cosh{(2r)}-\cos{(\varphi)}\sinh{(2r)})\omega 
\end{pmatrix},
\label{eq:singlemodestate}
\end{equation}
which is a covariance matrix of a squeezed thermal state.

In the multimode case where $M>1$ the state is essentially parameterized by $M$ thermal excitations, each independently and uniformly distributed, and $M$ squeezing parameters, also independently and uniformly distributed, in the following way. We begin from the product state of $M$ single mode thermal states, each with their own thermal excitations. Then we act with a random basis change, apply single mode squeezing of the position to all the modes with random magnitudes, and finally act on the resulting state with another random basis change. The random basis changes are built from Haar random $M\times M$ unitary matrices; let such a matrix be $\mathbf{U}$. Then by construction
\begin{equation}
    \mathbf{O}=\begin{pmatrix}
    \mathrm{Re}(\mathbf{U}) & \mathrm{Im}(\mathbf{U})\\
     -\mathrm{Im}(\mathbf{U}) & \mathrm{Re}(\mathbf{U})
    \end{pmatrix}
\end{equation}
is orthogonal and also a symplectic matrix w.r.t. the chosen ordering of operators.

For both STQM and channel equalization tasks we have chosen the intervals to be $n_{\mathrm{th}}\in[0,10]$ and $r\in[0,1]$. In the single mode case $\varphi\in[0,2\pi]$. These intervals also apply to the input states used for von Neumann entropy detection task shown in Fig.~\ref{fig:QCandCQ}\textbf{b}. In the entangler task the input states are always single mode vacuum states.

\subsection*{Training}

\subsubsection*{Cost function minimization}

A simple stochastic function optimizer called differential evolution (DE) is used. It treats the cost function as a black box, allowing it to, e.g., attempt to optimize functions where gradients (1st derivatives) or hessians (2nd derivatives) either do not exist or are not practical to calculate. Specifically, the implementation of Wolfram Mathematica 11.2 is used, which is described in Ref.~[\citeonline{wolfram}]. Here we give an overview of the method and the parameter values used; for full details consult the reference.

DE iterates a population of points $\{x_1,x_2,\ldots,x_d\}$. At each iteration a new population is created from the old one as follows. For each $x_j$ in the old population, three other old points $x_w$, $x_u$ and $x_v$ are chosen randomly and a point $x_s=x_w+s(x_u-x_v)$ is formed where $s\in\mathbb{R}$ is a parameter called scaling factor. Then a new point $x_j^{new}$ is created by taking each element either from $x_j$ or $x_s$ with probabilities $p$ and $1-p$, respectively, where the parameter $p$ is called cross probability. Finally, the new point $x_j^{new}$ replaces $x_j$ if $f(x_j^{new})$ is better than $f(x_j)$, where $f$ is a given cost or objective function. The stopping criterion is met when both $|f(x_j^{new})-f(x_j)|$ and $\lVert x_j^{new}-x_j\rVert$ are sufficiently small.

We initialize the population by generating $30 N M$ points, each corresponding to different interaction Hamiltonian $H_I$ where each interaction strength $g_{nm}$ between some reservoir oscillator $n\in\{1,\ldots,N\}$ and some input mode $m\in\{1,\ldots,M\}$ is uniformly and independently distributed in $g_{nm}\in[0,0.2]$ such that the spectral radius condition $\rho(\mathbf{A})\leq 0.99$ is satisfied. In the event that some point does not satisfy $\rho(\mathbf{A})\leq 0.99$ it is generated anew. 

We consistently use a scaling factor of $s=0.05$ and a cross probability $p=0.4$, i.e. rather small shifts are used to create the shifted points $x_s$ and when forming $x_j^{new}$ the elements are slightly more likely to be picked from $x_s$. We settled for these values through a simple lattice search. All other settings use the default values listed in Ref.~[\citeonline{wolfram}].

\subsubsection*{Cost function of the STQM task}

The cost function is
\begin{equation}
\begin{cases}
    f(H_I,\Delta t)=\lVert\mathbf{D}-\mathbf{I}\rVert+1/\lVert H_I\rVert_\infty& \textrm{if $\tau=0$,}\\
    f(H_I,\Delta t)=0.5\lVert\mathbf{D}\rVert+5\lVert\mathbf{C}\mathbf{A}^{\tau-1}\mathbf{B}-\mathbf{I}\rVert& \textrm{if $\tau>0$,}
\end{cases}
\label{eq:STQMobjective}
\end{equation}
where $\lVert\cdot\rVert$ is the Frobenius norm and where with a slight abuse of notation we have indicated by $\lVert H_I\rVert_\infty$ the maximum coupling strength between a reservoir oscillator and the input oscillator(s). The point of the term $1/\lVert H_I\rVert_\infty$ is to prevent the training to converge to the trivial solution $H_I=\mathbf{0}$. The factors $0.5$ and $5$ control the relative importance of minimizing the norm of $\mathbf{D}$ and achieving $\mathbf{C}\mathbf{A}^{\tau-1}\mathbf{B}\approx\mathbf{I}$; these values where chosen after some trial and error. While the function does not feature all of the relevant terms in Eqs.~\eqref{eq:STQMcondition}, numerical experiments suggest that including more terms leads to worse results.

\subsubsection*{Objective functions of the quantum channel equalization and entangler tasks}

Unlike in the relatively simple STQM task, in these tasks there is no obvious way to derive conditions on the reservoir symplectic matrix. This is why the objective function is the task dependent figure of merit---fidelity between reservoir output and original input in channel equalization and the logarithmic negativity in entangler---during training phase. 

\subsection*{Additional details about the quantum channel equalization task}

Let us write down the transformations caused by the channel and the reservoir at some timestep $k$. The interaction between input and channel modes induces a symplectic matrix $\mathbf{S}'$. Its action on all of the relevant modes reads
\begin{equation}
\begin{pmatrix}
\mathbf{x}^{Ch}_{k+1} \\
\mathbf{x}^R_{k} \\
\mathbf{x}^D_{k+1}
\end{pmatrix}=
\begin{pmatrix}
\mathbf{A}' & \mathbf{0} & \mathbf{B}' \\
\mathbf{0} & \mathbf{I} & \mathbf{0} \\
\mathbf{C}' & \mathbf{0} & \mathbf{D}' 
\end{pmatrix}
\begin{pmatrix}
\mathbf{x}^{Ch}_{k} \\
\mathbf{x}^R_{k} \\
\mathbf{x}^I_{k+1}
\end{pmatrix},
\end{equation}
where $\mathbf{S}'$ has already been divided into blocks such that $\mathbf{A}'$ is $C\times C$ and $\mathbf{D}'$ is $M\times M$. Nothing happens to the reservoir modes since there is no interaction between the reservoir and the channel. The reservoir processes $\mathbf{x^D_{k+1}}$ according to
\begin{equation}
\begin{pmatrix}
\mathbf{x}^{Ch}_{k+1} \\
\mathbf{x}^R_{k+1} \\
\mathbf{x}^O_{k+1}
\end{pmatrix}=
\begin{pmatrix}
\mathbf{I} & \mathbf{0} & \mathbf{0} \\
\mathbf{0} & \mathbf{A} & \mathbf{B} \\
\mathbf{0} & \mathbf{C} & \mathbf{D} 
\end{pmatrix}
\begin{pmatrix}
\mathbf{x}^{Ch}_{k+1} \\
\mathbf{x}^R_{k} \\
\mathbf{x}^D_{k+1}
\end{pmatrix}.
\end{equation}
Combining these two transformations we get
\begin{equation}
\begin{pmatrix}
\mathbf{x}^{Ch}_{k+1} \\
\mathbf{x}^R_{k+1} \\
\mathbf{x}^O_{k+1}
\end{pmatrix}=
\begin{pmatrix}
\mathbf{A}' & \mathbf{0} & \mathbf{B}' \\
\mathbf{BC}' & \mathbf{A} & \mathbf{BD}' \\
\mathbf{DC}' & \mathbf{C} & \mathbf{DD}' 
\end{pmatrix}
\begin{pmatrix}
\mathbf{x}^{Ch}_{k} \\
\mathbf{x}^R_{k} \\
\mathbf{x}^I_{k+1}
\end{pmatrix},
\label{eq:Qchanstep}
\end{equation}
where the intermediate form $\mathbf{x^D}$ of the input modes has been eliminated. The dynamics now follows Eqs.~\eqref{eq:sympmatstep} through \eqref{eq:QQinputdynamics} with the replacements
\begin{equation}
\mathbf{x}_k^R\mapsto\mathbf{x}_k^{Ch}\oplus\mathbf{x}_k^R,\quad
    \mathbf{A}\mapsto\begin{pmatrix}
\mathbf{A}' & \mathbf{0} \\
\mathbf{BC}' & \mathbf{A} 
\end{pmatrix},\quad
    \mathbf{B}\mapsto\begin{pmatrix}
\mathbf{B}' \\
\mathbf{BD}' 
\end{pmatrix},\quad
    \mathbf{C}\mapsto\begin{pmatrix}
\mathbf{DC}'& \mathbf{C} 
\end{pmatrix},\quad
     \mathbf{D}\mapsto\mathbf{DD}',
\label{eq:Qchanreplace}
\end{equation}
that is to say the channel and the reservoir may be treated together as if they formed a new, larger reservoir. This simplifies the equations of motion and the simulation of the dynamics. Although one may now consider Eqs.~\eqref{eq:STQMcondition} to solve the task, in practice the performance is very poor because only the reservoir blocks are controllable, hence the modifications of Eqs.~\eqref{eq:Qchantask}.

\subsection*{Additional details about the von Neumann entropy detection task}

Let $\rho$ be a single mode Gaussian state. Then its von Neumann entropy is defined as $S_V(\rho)=-\mathrm{Tr}(\rho\mathrm{ln}(\rho))$. It can be shown\cite{agarwal1971entropy} that
\begin{equation}
  S_V(\rho)=n_{\mathrm{th}}\mathrm{ln}\left(\frac{n_{\mathrm{th}}+1}{n_{\mathrm{th}}}\right)+\mathrm{ln}(n_{\mathrm{th}}+1)
  \label{eq:vonneumann}
\end{equation}
where $n_{\mathrm{th}}$ is the amount of thermal excitations of the state $\rho$. This quantity in turn is connected to the determinant of the associated covariance matrix $\sigma$ through
\begin{equation}
    \mathrm{Det}(\sigma)=(0.5+n_{\mathrm{th}})^2,
    \label{eq:determinant}
\end{equation}
which can be seen by direct calculation starting from, e.g., Eq.~\eqref{eq:singlemodestate}. In Fig.~\ref{fig:QCandCQ}\textbf{b} the actual $S_V(\rho)$ is compared to that computed from the estimated determinant of the input covariance matrix using Eqs.~\eqref{eq:vonneumann} and \eqref{eq:determinant}.

\section*{Data availability}

Data is available from the corresponding author upon reasonable request.

\bibliography{references}

\section*{Acknowledgements}

The author acknowledges the Spanish State Research Agency, through the Severo Ochoa and Mar\'ia de Maeztu Program for Centers and Units of Excellence in R\&D (MDM-2017-0711) and through the  QUARESC project (PID2019-109094GB-C21 and -C22/ AEI / 10.13039/501100011033).  The author also acknowledges funding by CAIB through the QUAREC project (PRD2018/47). The author acknowledges support from the Turku Collegium for Science, Medicine and Technology. Finally, the author would like to thank Roberta Zambrini, Gian Luca Giorgi and Miguel C. Soriano for helpful discussion and comments. 

\section*{Author contributions statement}

J.N. designed the research project, carried out the analytical calculations and the numerical experiments, analysed the results and wrote the manuscript.

\section*{Additional information}

The author declares no competing interests.

\end{document}